\documentclass{article}
\pdfoutput=1 
\usepackage{jheppub}
\usepackage{graphicx} %
\usepackage{amsmath,amssymb,commath,braket}
\usepackage[most]{tcolorbox}
\usepackage{mathtools}
\usepackage{cleveref}
\usepackage{bbm}
\usepackage{caption}
\usepackage{subcaption}
\usepackage{soul}
\usepackage{cancel}
\usepackage{orcidlink}
\usepackage{placeins}

\usepackage[T1]{fontenc} 

\usepackage{xcolor}

\newcommand{\sSI}{\sigma_{\rm SI}}

\usepackage{braket}

\title{Heavy Inert Doublet Reconciles LZ and IceCube}

\author[a,b,c,1]{Hitoshi Murayama\,\orcidlink{0000-0001-5769-9471},\note{Hamamatsu Professor}}\emailAdd{hitoshi@berkeley.edu}
\author[a,b]{Bea Noether\, \orcidlink{0000-0002-2947-3210},}\emailAdd{bea\_noether@berkeley.edu}

\affiliation[a]{Leinweber Institute for Theoretical Physics, University of California, Berkeley, CA 94720, USA}
\affiliation[b]{Kavli Institute for the Physics and Mathematics of the Universe (WPI), University of Tokyo, Kashiwa 277-8583, Japan}
\affiliation[c]{Ernest Orlando Lawrence Berkeley National Laboratory, Berkeley, CA 94720, USA}

\abstract{
We show that the inert doublet dark matter can reconcile the purported LZ direct detection signal and IceCube constraints. The preferred dark matter mass is 11.3--20.3~TeV, an order of magnitude heavier than the standard higgsino target. The smaller number density at a high mass can reproduce the LZ event given the exponential sensitivity to the velocity, while suppressing the capture rate in the Sun because of the higher infall velocity. On the other hand, the high mass requires a large coupling between the inert doublet and Higgs boson to have the correct annihilation cross section for the freeze-out. There is about an order of magnitude of validity range below the Landau pole, beyond which a UV completion is needed. Electroweak precision observables are below the experimental limits despite the large coupling. }

\begin{document}
\maketitle
\flushbottom

\section{Introduction}
\label{sec:intro}

Dark matter dominates the matter density of the universe today and played a crucial role in structure formation. However, its nature is not known beyond its gravitational impacts. An enormous amount of effort has been devoted to its experimental detection. Direct detection experiments have improved the sensitivity over the decades by more than six orders of magnitude and are approaching the level where even neutrinos can constitute the background. Among them, those based on liquid xenon with a two-phase time projection chamber are leading the field: LZ \cite{LZ:2024zvo}, XENONnT \cite{XENON:2025vwd}, and PandaX-4T \cite{PandaX:2024qfu}. 

Recently, the LZ experiment reported a tantalizing event that is unlikely to be explained as a background event \cite{LZ:2026axp}. On the other hand, the energy deposit is unusually high, about 250~keV if interpreted as nuclear recoil. Given the lack of events with lower energies, many papers appeared that interpret the event as a result of an inelastic upscattering event \cite{Tucker-Smith:2001myb}. A prototypical example in a well-motivated model is the 1.1~TeV higgsino in supersymmetry \cite{Fan:2026kxx,Freese:2026sga,Wu:2026nhi}. Yet it was also pointed out that such a higgsino would be captured by the sun, accumulate in the core, and annihilate into high-energy neutrinos whose flux would exceed the constraints from the IceCube experiment \cite{Pospelov:2026ewn,DiMauro:2026dqp,Bose:2026ndd,Nguyen:2026lui,Ghosh:2026txe}. Non-thermal production of a heavy higgsino may evade this problem \cite{Langhoff:2026ujr}. There is also a potential issue with the lack of events in the higher energy bin \cite{Rodd:2026tyn}.

We point out that the purported LZ event can be reconciled with the IceCube constraint if the dark matter is not a higgsino but rather the ``inert doublet'' \cite{Cirelli:2005uq,Barbieri:2006dq}. Namely that the dark matter is a neutral component of an electroweak-doublet scalar $\Phi$ with no vacuum expectation value or Yukawa coupling to quarks and leptons. It is assigned a ${\mathbb Z}_2$ odd parity $\Phi \rightarrow -\Phi$ and hence stable. It is supposed be a Weakly Interacting Massive Particle (WIMP), namely that its abundance as a thermal relic is fixed by its annihilation cross section at the freeze-out. The coupling to the standard model Higgs splits it to four states, $\phi_i (i=0,1,2,3)$,
\begin{align}
    \Phi 
    = \left( \begin{array}{cc} \Phi^+ \\ \Phi^0 \end{array} \right)
    = \left( \begin{array}{cc} \phi^+ \\ \frac{\phi_0 + i \phi_3}{\sqrt{2}} \end{array} \right)
\end{align}
where $\phi^\pm = \frac{1}{\sqrt{2}}(\phi_1\pm i \phi_2)$. For concreteness, let us assume $\phi_0$ is dark matter, with $\phi_3$ only $\sim 370$~keV heavier. In this case, the inelastic upscattering is $\phi_0 {\rm Xe} \rightarrow \phi_3 {\rm Xe}$ due to the $Z$-exchange. 

The important point in our proposal is to consider a very high mass for the inert doublet dark matter in the 10--20~TeV range. Recall the original papers assumed $m_{\phi_0} = 540$~GeV \cite{Cirelli:2005uq} or 70~GeV \cite{Barbieri:2006dq}. Naively it appears our mass range is too heavy for a WIMP for its annihilation cross section by gauge interactions. However the annihilation of the inert doublet dark matter is dominated by the longitudinal polarizations $W_L^+ W_L^-$ and $Z_L Z_L$ in the final state if it is much heavier than $m_W$ \cite{Barbieri:2006dq,Treesukrat:2024akz}. In this case the annihilation cross section is determined not by the gauge couplings but rather a coupling in the scalar potential, which can be understood with the equivalence theorem that identifies the longitudinal $W_L$, $Z_L$ as the Nambu--Goldstone bosons in the standard model Higgs doublet in the high-energy limit \cite{Lee:1977eg,Gounaris:1986cr}. The coupling is arbitrary and may be large, allowing for a much larger annihilation cross section and hence heavier WIMP dark matter. This possibility was made especially clear in \cite{Treesukrat:2024akz}.

The possibility of a doublet scalar dark matter was discussed in several papers to explain the LZ event. Refs.~\cite{Nomura:2026qyq,Wang:2026ytg,Bandyopadhyay:2026gjw} did not address the IceCube constraint, and \cite{Bandyopadhyay:2026gjw} introduced an extra singlet scalar. Ref.~\cite{Lian:2026hpm} added six extra scalars in the context of sneutrinos and tried to evade the IceCube constraint with different annihilation channels. Ref.~\cite{Arcadi:2026kev} added an axion portal. All of them focused on dark matter mass at order TeV and below. Our proposal is about a much heavier dark matter in the 11--20~TeV range as mentioned above and is qualitatively different from all of them.

This paper is organized as follows. We introduce the inert doublet dark matter mode and our notation in Section~\ref{sec:inert}. The LZ event is studied in Section~\ref{sec:lzsignal} so that it is interpreted as an inelastic upscattering of dark matter. We determine the range of dark matter mass and its mass splitting. In Section~\ref{sec:solar}, we study constraints on the high-energy neutrino from the IceCube experiment and further constrain the parameter space. We find that a heavy inert doublet of 11--20~TeV can reconcile the LZ event and the IceCube constraint. It requires a relatively large coupling $\gtrsim 7$. We examine the internal consistency of this large coupling in Section~\ref{sec:consistency} and find the parameter region of consistency. We conclude in Section~\ref{sec:conclusion}.

\section{Inert Doublet Dark Matter}
\label{sec:inert}

\subsection{The Model}
The inert doublet $\Phi$ has the same electroweak quantum numbers as the standard model Higgs doublet $H$, namely a doublet under $SU(2)_L$ with hypercharge $Y=+1/2$. We assign a new ${\mathbb Z}_2$ parity under which $\Phi \rightarrow -\Phi$ while all other particles are even. We assume $\Phi$ has no VEV or Yukawa couplings to quarks and leptons. We use the notation
\begin{align}
    H = \left( \begin{array}{c} H^+ \\ H^0 \end{array} \right)
    = \left( \begin{array}{c} \chi^+ \\ \frac{v+h+i\chi^0}{\sqrt{2}} \end{array} \right),
\end{align}
where $v=246$~GeV, $h$ is the 125~GeV Higgs boson, while $\chi^+$ and $\chi^0$ are eaten by $W^+$ and $Z$ bosons, respectively. 

The most general potential of $H$ and $\Phi$ is \cite{Treesukrat:2024akz}
\begin{align}
    V=&
    m_H^2 H^\dagger H +\frac{1}{2} \lambda_H (H^\dagger H)^2
    + M^2 \Phi^\dagger \Phi +\frac{1}{2} \lambda_\Phi (\Phi^\dagger \Phi)^2 \nonumber \\
    & + \eta (H^\dagger H)(\Phi^\dagger \Phi)
    + \kappa (H \Phi)(H \Phi)^*
    - \frac{\zeta}{2}
    \left[(H^\dagger\Phi)^2+\mathrm{h.c.}\right],
\end{align}
where we introduced our notation $m_H^2 = -\mu_1^2$, $M^2 = \mu_2^2$, $\lambda_H \equiv \lambda_1, \lambda_\Phi \equiv \lambda_2, \eta\equiv \lambda_3+\lambda_4$, $\kappa\equiv -\lambda_4$, and $\zeta \equiv -\lambda_5$, compared to those in \cite{Treesukrat:2024akz}. We used the identity $ (H^\dagger \Phi)(\Phi^\dagger H) = (H^\dagger H)(\Phi^\dagger \Phi) - (H \Phi)(H \Phi)^*$, where $(H \Phi) = H^+ \Phi^0 - H^0 \Phi^+$. Note that there is a global $U(1)_\Phi$ symmetry that rotates $\Phi \rightarrow e^{i\theta} \Phi$ in the absence of $\zeta$ which breaks $U(1)_\Phi$ to ${\mathbb Z}_2$. Therefore it is technically natural for $\zeta \ll 1$ to be small. The $U(1)_\Phi$ rotation can always make $\zeta$ real and positive, and hence $\phi_0$ is the dark matter. Note that $M$ can be as high as 80~TeV within the unitarity limit \cite{Treesukrat:2024akz}.

These relate to physical masses as
\begin{align}
    m_{\phi^\pm}^2 =& M^2 + \frac{1}{2}(\kappa+\eta) v^2,
    &
    m_{\phi_0}^2 =& M^2 + \frac{1}{2}(\eta-\zeta)v^2,
    &
    m_{\phi_3}^2 =& M^2 + \frac{1}{2}(\eta+\zeta)v^2.
\end{align}
Note that technically natural $\zeta \ll 1$ keeps mass splitting between $\phi_0$ and $\phi_3$ small.  In addition, we will assume $M \gg v$, a rather large $\kappa \approx 10$, while a small $\eta\approx 10^{-2}$  when we discuss phenomenology. 

Define the mass-splittings:
\begin{align}
    \delta_0 \equiv&~ m_{\phi_3}-m_{\phi_0} \approx \frac{\zeta v^2}{2M},
    &
    \Delta_\pm \equiv&~ m_{\phi^\pm} - m_{\phi_0} \approx \frac{(\kappa+\zeta) v^2}{4M}.
\end{align}
The gauge kinetic term contains the off-diagonal $Z$ interaction for inelastic scattering:
\begin{align}
    \mathcal{L} \supset&~ \frac{1}{2} g_Z Z_\mu (\phi_3\partial^\mu \phi_0 - \phi_0\partial^\mu \phi_3),
    \label{eq:ZAH_offDiagonal}
\end{align}
where $g_Z = e/s_W c_W = g/c_W$. 

\subsection{Thermal freeze-out}

All four states of $\Phi$ are nearly degenerate at the time of the freeze-out and co-annihilate. To see this, we estimate the mass splittings. For the neutral splitting suggested by inelastic direct detection, the required mass splitting is basically the same as the higgsino case, see Section~\ref{sec:lzsignal}. It constrains the coupling $\zeta$,
\begin{align}
    \delta_0 &= m_{\phi_3} - m_{\phi_0}= {350~{\rm keV}} \left(\frac{\zeta}{1.62\times10^{-4}}\right)
    \left(\frac{14~{\rm TeV}}{m_{\phi_0}}\right)
\end{align}
For the charged splitting, we consider a large $\kappa$,
\begin{align}
    \delta_\pm &= m_{\phi^\pm} - m_{\phi_0} = {60~{\rm GeV}} \left(\frac{\kappa}{10}\right)
    \left(\frac{14~{\rm TeV}}{m_{\phi_0}}\right).
\end{align}
Both mass splittings are much smaller than the freeze-out temperature $T_F = m_{\phi_0}/x_f\approx 520$~GeV for $m_{\phi_0}=14$~TeV and $x_f \approx 27$, and therefore effectively degenerate during freeze-out, and $\zeta$ and gauge interactions have a negligible effect on their annihilation rates. This allows the direct-detection splitting to be treated independently, to an excellent approximation, from the calculation of the thermal abundance. In fact, $T_f \gg v$ and hence the electroweak symmetry is effectively unbroken $v=0$ at that time. Therefore, we can use electroweak eigenstates $\Phi^0, \Phi^+, H^+, H^0$. We ignore $m_H \ll M$ below.

The annihilation cross sections at the threshold are dominated by the coupling $\kappa$, 
\begin{align}
    \sigma v (\Phi^0 \Phi^{0*} \rightarrow H^+ H^-)
    &= \sigma v (\Phi^+ \Phi^{+*} \rightarrow H^0 H^{0*})
    = \sigma v (\Phi^0 \Phi^{+*} \rightarrow H^0 H^{+*}) \nonumber \\
    & = \sigma v (\Phi^+ \Phi^{0*} \rightarrow H^+ H^{0*})
    = \frac{\kappa^2}{32\pi M^2}.
    \label{eq:annihilation}
\end{align}
They are consistent with \cite{Treesukrat:2024akz} where cross sections are computed in the broken phase once recalling the Goldstone-boson equivalence theorem \cite{Lee:1977eg,Gounaris:1986cr}: the longitudinal modes $W_L$ and $Z_L$ can be identified with the eaten Goldstone boson states $\chi^\pm$ and $\chi^0$ in $H$.
$\Phi^0$ can be annihilated by either $\Phi^{0*}$ or $\Phi^{+*}$ but not $\Phi^0$ or $\Phi^+$, a chance encounter of $2/4=1/2$, and hence the overall annihilation cross section is
\begin{align}
    \langle\sigma_{\rm eff} v \rangle
    &= \frac{1}{2} \frac{\kappa^2}{32\pi M^2}.
\end{align}

As we will see in Sections~\ref{sec:lzsignal} and \ref{sec:solar}, the LZ event and the IceCube constraint can be reconciled for the mass range $M=11.3$--20.3~TeV. We have run Boltzmann equations with $g_*=106.75$, $\Omega_c h^2=0.1200$ to find
\begin{align}
    \langle \sigma_{\rm eff} v \rangle = 2.17\times 10^{-26}~{\rm cm}^3{\rm s}^{-1},
\end{align}
to obtain the correct abundance. Therefore we need
\begin{align}
    \kappa &\approx 6.92 \left( \frac{M}{11.3~{\rm TeV}}\right).
    \label{eq:kappaM}
\end{align}
We use full point-wise solved values of $\kappa$ for the actual numerics, but the above proportionality is a good approximation at a percent level.

\begin{figure}[t]
    \centering
    \includegraphics[width=0.7\textwidth]{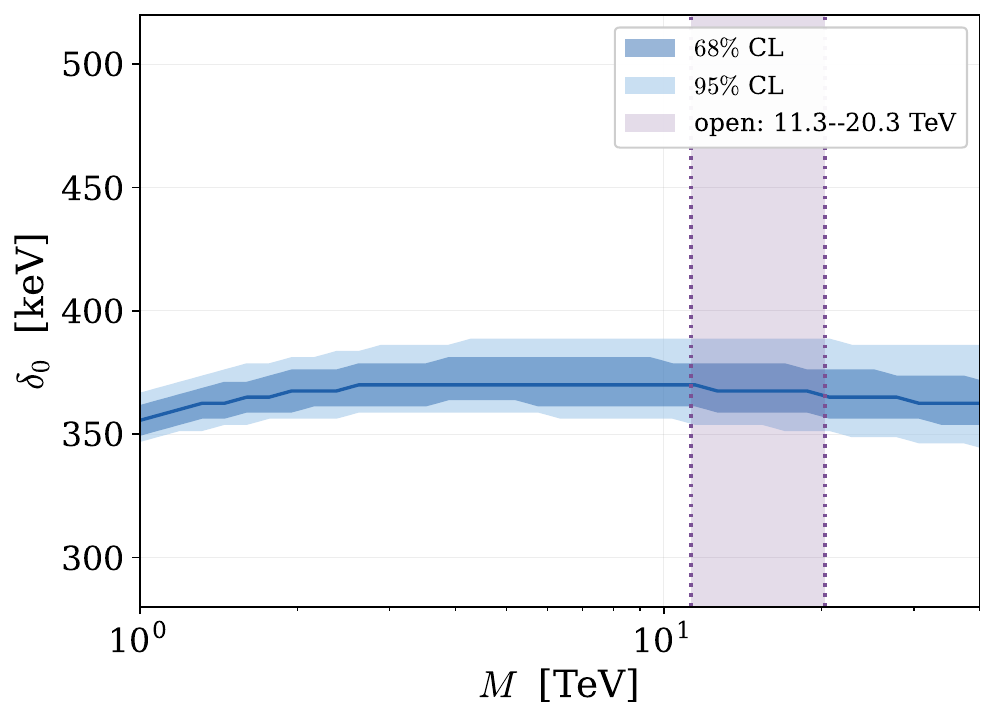}
    \caption{Likelihood for the inert-doublet interpretation of the LZ 248~keV recoil, in the
plane of the dark matter mass $M$ and the splitting $\delta_0$ between the two neutral
states. Shaded: the joint 68\% and 95\% regions, $\Delta(-2\log\mathcal{L}) = 2.30$ and
5.99 for two degrees of freedom, relative to the global maximum. The solid curve is the
maximum-likelihood $\delta_0$ at each mass. The likelihood is built from the recoil
spectrum alone, with the standard halo model and a Helm form factor. The purple band is
the mass range the model leaves open once solar limits and perturbative unitarity are
enforced: bounded below at $11.3\,$TeV, where the LZ $1\sigma$ band in $\delta_0$ stops
being excluded by the solar-neutrino limit \cite{IceCube:2025fcu} at
$\sigma_{\rm SI}^{\rm loop}=0$ (see discussions in Section~\ref{sec:solar}), and above at $20.3\,$TeV by tree-level perturbative
unitarity, $\kappa<4\pi$, with the thermal relic density imposed.}
    \label{fig:lzcontour}
\end{figure}

\section{LZ Signal}
\label{sec:lzsignal}

The off-diagonal $Z$ coupling of Eq.~\eqref{eq:ZAH_offDiagonal} is identical in structure to the
split-higgsino's off-diagonal vector coupling to the two nearly-degenerate
Majorana mass eigenstates~\cite{Fan:2026kxx}, up to the trivial replacement of a
real scalar pair by a pseudo-Dirac fermion pair. Explicit evaluation of the two
nuclear matrix elements gives
\begin{align}
    |\mathcal{M}_{\rm scalar}|^2 &= 4a^2, &
    |\mathcal{M}_{\rm fermion}|^2 &= 4a^2 - (4m_N^2-t)(\delta_0^2-t),
\end{align}
with $a\equiv p\cdot(k+k')$, $t\equiv-2m_NE_R$, and $p,k,k'$ the incoming dark
matter, incoming nucleus, and outgoing nucleus four-momenta. The two differ
only at $\mathcal{O}(m_NE_R/M^2)\sim10^{-9}$ for $M\sim\mathcal{O}(1\text{--}10)$~TeV,
so the inert doublet inherits the higgsino's direct-detection phenomenology
essentially unchanged: the same nucleon-level cross section
$\sigma_n = G_F^2\mu_n^2/2\pi \simeq 7.4\times10^{-39}~{\rm cm^2}$, independent
of $M$ and $\delta_0$ for $M\gg m_N$.

\begin{figure}[t]
    \centering
    \includegraphics[width=0.7\textwidth]{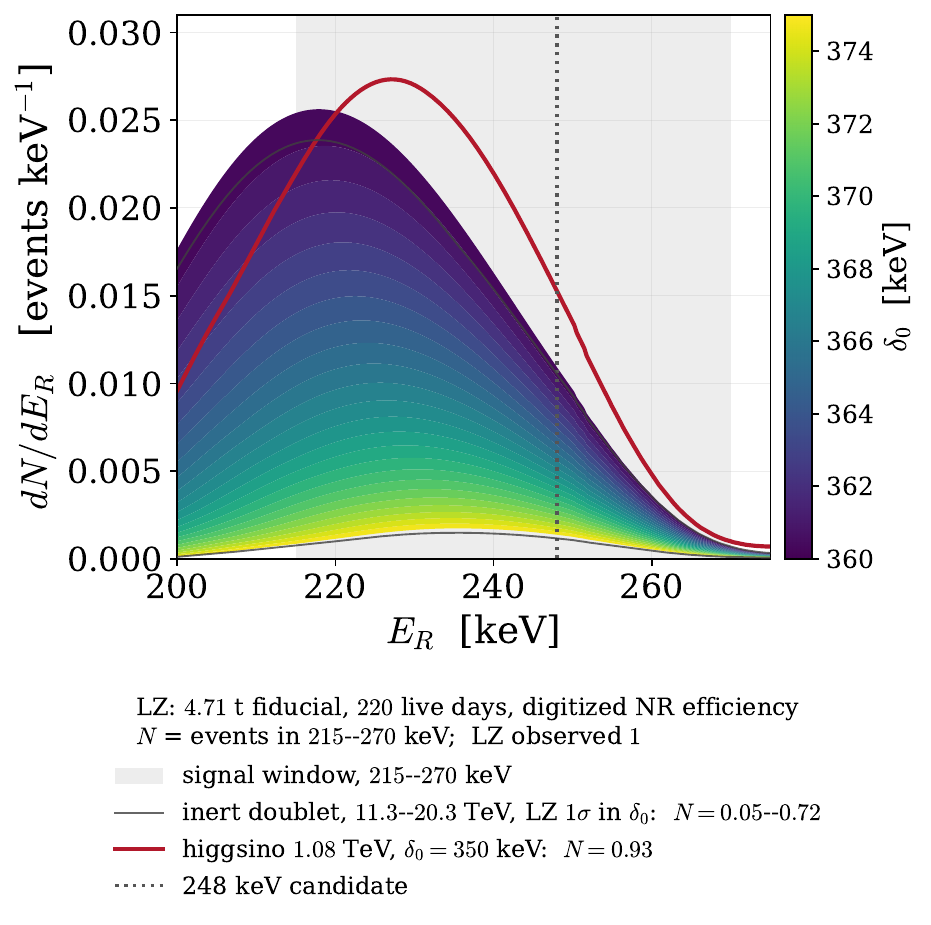}
    \caption{Detector-level recoil spectrum of the inert doublet in LZ, $dN/dE_R =
\mathcal{E}\times\epsilon(E_R)\times dR/dE_R$, using the published exposure
(4.71~t fiducial $\times$ 220 live days) and the digitised nuclear-recoil efficiency. The
colour gradient sweeps the splitting $\delta_0$ across its LZ $1\sigma$ range,
$362$--$375\,$keV, at a single representative mass --- the midpoint of the open window,
$M = 15.8\,$TeV. The two thin grey curves outline the envelope over the whole allowed
region, every mass in $M = 11.3$--$20.3\,$TeV combined with the full $1\sigma$ range in
$\delta_0$ at each; they sit only slightly outside the gradient, because across this
window the mass dependence of the rate is a $\sim10\%$ effect while $\delta_0$ moves it by
more than an order of magnitude. The grey band is the $215$--$270\,$keV signal window and
the dotted vertical line the candidate recoil at $E_R = 248\,$keV. Integrated over the
window the inert doublet predicts $N = 0.05$--$0.72$ events, against $N = 0.93$ for the
$1.08\,$TeV higgsino benchmark at $\delta_0 = 350\,$keV (red); LZ observed one event.}
    \label{fig:lzspectrum}
\end{figure}

We evaluate the extended likelihood of Ref.~\cite{Fan:2026kxx} directly for $H$,
using their Standard Halo Model ($v_0=220$, $v_{\rm esc}=540$~km/s,
$\rho_H=0.4~{\rm GeV/cm^3}$), Earth's velocity averaged over one year, and the
full LZ exposure and recoil window ($2.84$~tonne-yr, $5.4$--$269.9$~keV)
containing the candidate event at $E_R=248\pm23({\rm stat})\pm23({\rm sys})$~keV.
We use the Helm form factor, which Ref.~\cite{Fan:2026kxx} finds changes the
preferred splitting by only $5$--$10$~keV relative to the shell-model form
factor of Ref.~\cite{Vietze:2014vsa}.

Scanning $M\in[0.3,100]$~TeV and $\delta_0\in[150,500]$~keV, the global
best fit is $M=2.1$~TeV, $\delta_0=368$~keV, with the nominal
$\Delta(-2\log\mathcal{L})=2.30$ region extending, essentially unbroken, across
the entire scanned mass range (Fig.~\ref{fig:lzcontour}) -- the number density
suppression $\propto1/M$ is compensated by the halo integral's growth as
$v_{\min}$ falls with increasing $M$~\cite{Fan:2026kxx}. The LZ event therefore
does not by itself select a mass, and we do not claim that it does: the window
$M=11.3$--$20.3$~TeV quoted throughout is set from below by the solar-neutrino
limit (Section~\ref{sec:solar}) and from above by perturbative unitarity
(Section~\ref{sec:consistency}). What LZ does determine, and determines tightly,
is the splitting -- and it determines it almost independently of the mass. Across
the whole window the maximum-likelihood splitting is constant to within the
$2.5$~keV resolution of our scan, and the $68\%$ interval is likewise essentially
mass-independent,
\begin{equation}
    \delta_0 = 368^{+12}_{-8}~{\rm keV}
    \qquad (11.3~{\rm TeV} < M < 20.3~{\rm TeV}),
    \label{eq:delta0band}
\end{equation}
quoted at $\Delta(-2\log\mathcal{L})=2.30$ for the two parameters $(M,\delta_0)$,
as in Fig.~\ref{fig:lzcontour}. No point in the band lies more than about one
unit of $-2\log\mathcal{L}$ from the global best fit. Integrated over the
$215$--$270$~keV signal window, the expected yield along the best-fit ridge is
$N=0.22$--$0.33$ events across the band; sweeping $\delta_0$ over its profiled
$1\sigma$ range at each mass ($\Delta(-2\log\mathcal{L})=1$, the range displayed
in Fig.~\ref{fig:lzspectrum}) spreads this to $N=0.05$--$0.72$, against $N=0.93$
for the $1.08$~TeV higgsino benchmark in the same window. Over the full
$5.4$--$269.9$~keV analysis window the corresponding numbers are
$N=0.27$--$0.43$, $N=0.06$--$1.25$ and $N=1.23$. LZ reports one event. A heavy
inert doublet anywhere in the $11.3$--$20.3$~TeV window can therefore account for
the LZ candidate, with no adjustment beyond the value of $\delta_0$ already
preferred by Fig.~\ref{fig:lzcontour}.

\section{Solar Constraints}
\label{sec:solar}

Ref.~\cite{Bose:2026ndd} derived the strongest bound on the higgsino interpretation of the LZ event. They find that below 11.3~TeV the dark matter is captured efficiently enough in the Sun and annihilates into a large enough flux of neutrinos to be excluded by IceCube \cite{IceCube:2025fcu}, unless the elastic spin-independent cross section on nucleons $\sSI$ is small enough to suppress capture. They note that for $\sSI\gtrsim 10^{-48}~{\rm cm}^2$ capture is already equilibrium-limited and the constraints saturate.

We re-implemented the solar-capture-to-neutrino-flux calculation of
Ref.~\cite{Bose:2026ndd} ourselves, following their equations in order: the capture rate
as a function of $(M,\sigma_{\rm SI},\sigma_{\rm SD})$ (their Eq.~(1), whose formalism
follows Ref.~\cite{Garani_2017}), the dark-matter orbit's shrinkage toward full
thermalization inside the Sun (their Eq.~(5), following Ref.~\cite{Pospelov:2026ewn}), the
resulting annihilation rate (their Eqs.~(4) and~(6)), and the $\chi^2$ comparison to the
null neutrino observation (their Eq.~(8)), with the $95\%$ C.L. boundary at
$\Delta\chi^2 = 2.71$ for one degree of freedom. In our case, the annihilation proceeds via the cross section $\sigma(\phi_0 \phi_0 \rightarrow W_L^+ W_L^-) = \frac{1}{2} \sigma( \Phi^0 \Phi^{0*} \rightarrow H^+ H^-)$ in Eq.~\eqref{eq:annihilation} thanks to the equivalence theorem.

The inputs are as follows. Solar density and composition profiles are taken from the
BS05(AGS,OP) model of Ref.~\cite{Bahcall_2005}, using its own tabulated mass fractions
for $^{1}$H, $^{4}$He, $^{3}$He, $^{12}$C, $^{14}$N and $^{16}$O. Neutrino spectra are
generated with \texttt{$\chi$aro$\nu$}~\cite{Liu_2020}, which includes decays and interactions
in the solar interior together with electroweak corrections, and propagated from the Sun
to $1\,$AU with \texttt{nuSQuIDS}~\cite{Arg_elles_2022}, including oscillations, charged- and
neutral-current interactions and $\tau$ regeneration. For the IceCube $\nu_\mu$ channel we
use the effective area $A_{\rm eff}(E_\nu)$ and median angular resolution
$\Delta\theta(E_\nu)$ entering their Eqs.~(9)--(10), together with the observed and
atmospheric-background event distributions in $\cos\psi$ entering Eq.~(8), all from
Ref.~\cite{IceCube_2016}.

\begin{figure}[tbh]
\centering
\includegraphics[width=\linewidth]{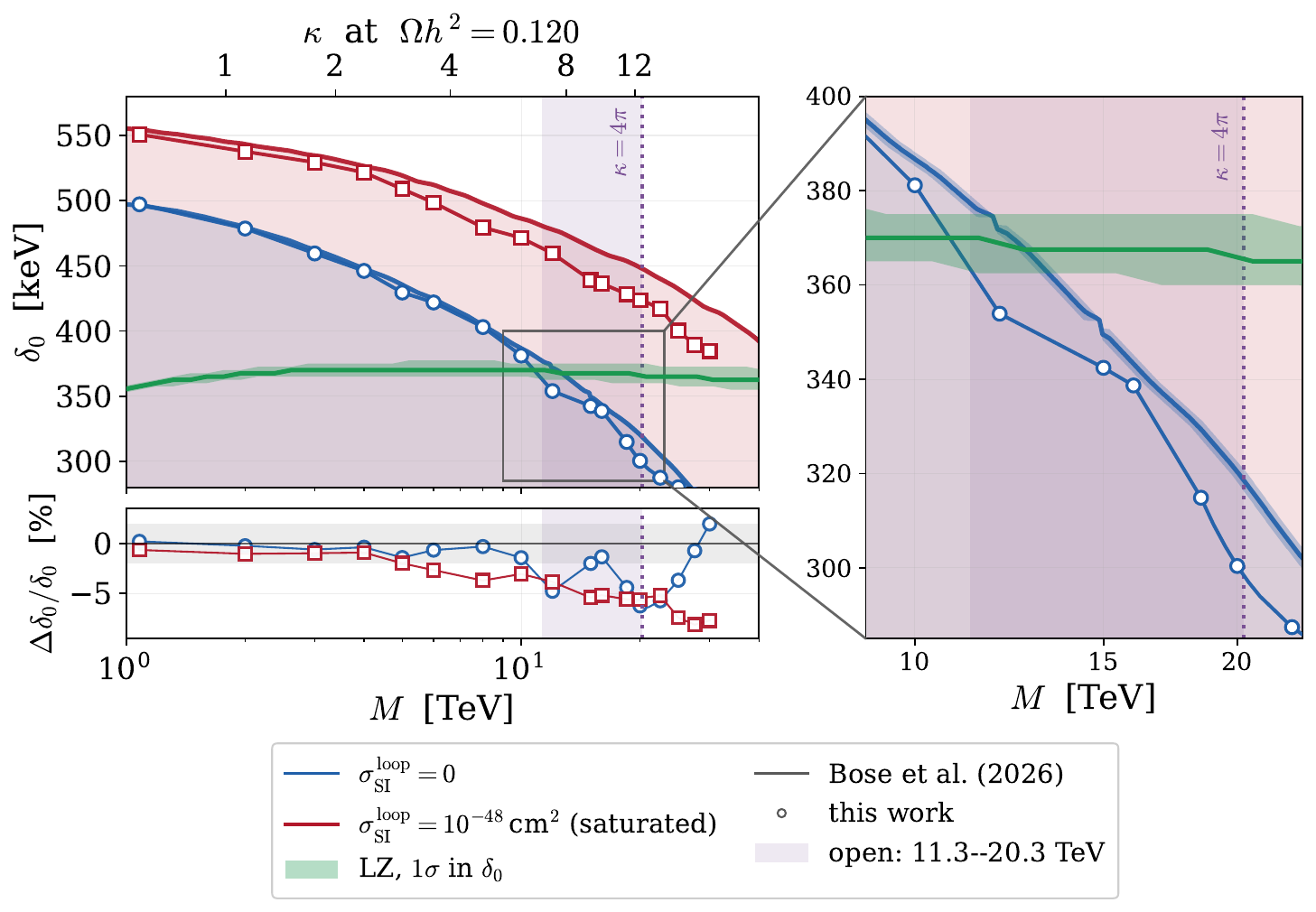}
\caption{Solar-neutrino limit on the heavy inert doublet in the $(M,\delta_0)$ plane.
Regions below the curves are excluded. Blue and red curves are the two loop-induced
elastic cross sections bracketed in Ref.~\cite{Bose:2026ndd},
$\sigma_{\rm SI}^{\rm loop}=0$ and $10^{-48}\,$cm$^2$ (saturated); solid curves are our
digitisation of their IceCube $\nu_\mu$ limit (linewidth shaded), open markers our own
independent calculation at the model's candidate masses. We set
$\eta\equiv\lambda_3+\lambda_4=0$, so the tree-level elastic cross section vanishes. The
top axis shows the coupling $\kappa\equiv-\lambda_4$ required for thermal freeze-out at
$\Omega h^2=0.120$, bounded by perturbative unitarity ($\kappa<4\pi$, dotted line at
$M=20.3\,$TeV). The shaded vertical band is the window the model leaves open,
$M = 11.3$--$20.3\,$TeV: bounded below where the LZ $1\sigma$ band in $\delta_0$ clears
the $\sigma_{\rm SI}^{\rm loop}=0$ boundary, and above by that unitarity ceiling. The
green band is the LZ $1\sigma$ range in $\delta_0$. Right: magnification of the open
window. Lower panel: fractional difference between our computed boundaries and the
digitised ones (grey band: $\pm2\%$).}
\label{fig:Mdelta}
\end{figure}

As a check, we reproduced their digitized Fig.~3 boundaries: our independent calculation
agrees with the IceCube $\nu_\mu$ channel~\cite{IceCube_2016} --- the strongest constraint
over this mass range, and the one we use throughout --- to within $2\%$ over
$1.08$--$5\,$TeV in both panels. Across the open window the agreement is $1$--$6\%$, our
computed boundary lying below the digitized one in both panels (lower panel of
Fig.~\ref{fig:Mdelta}); the largest excursion anywhere in the window is $-6.4\%$, near
$20.5\,$TeV. Throughout the window, and indeed out to $\sim28\,$TeV, the difference is in
the conservative direction --- a weaker computed limit than the digitized one --- so
nothing we claim below rests on the residual. Only above $\sim28\,$TeV, far outside the
region of interest, does the $\sigma_{\rm SI}^{\rm loop}=0$ residual change sign and our
boundary become marginally the stronger of the two.

One approximation should be recorded. The propagated spectra were generated at
$m_\chi = 1080\,$GeV --- the higgsino benchmark against which we validate --- and the same
per-annihilation yield is used at every mass; the capture, thermalization and annihilation
chain is evaluated at the true mass throughout, so the approximation enters only through
the shape of the neutrino spectrum. Its effect is bounded and small. Absorption in the Sun
removes essentially all of the high-energy flux before it reaches $1\,$AU: at the
benchmark, the surviving $\nu_\mu$ spectrum is attenuated by a factor $27$ at $400\,$GeV
and $500$ at $1\,$TeV, so that $99\%$ of the flux at $1\,$AU lies below $\simeq500\,$GeV
and under $0.5\%$ above $600\,$GeV. The detectable signal is therefore built from a region
of the spectrum well below the endpoint, and a heavier dark matter particle changes it
only through the hardness of the injected spectrum below that cutoff. The residual quoted
above bounds the size of that effect directly: with
$d\ln\Gamma_{\rm ann}/d\ln\delta_0 \simeq -9$ near the boundary, the $1$--$6\%$ agreement
in $\delta_0$ corresponds to a per-annihilation yield differing by no more than a factor
$1.1$--$1.7$. 

This calculation is what fixes the lower edge of the viable mass range. In
Fig.~\ref{fig:Mdelta} the $\sigma_{\rm SI}^{\rm loop}=0$ boundary falls with increasing
mass and crosses the LZ-preferred band in $\delta_0$ near $11\,$TeV: below the crossing the
splitting LZ wants is excluded by the solar-neutrino limit, above it the splitting is
allowed. Requiring the \emph{entire} LZ $1\sigma$ band in $\delta_0$ to clear the boundary
gives $M > 11.3\,$TeV, and it is this conservative choice we quote. Together with the
unitarity ceiling of Section~\ref{sec:consistency} this brackets the model into
$11.3\,{\rm TeV} < M < 20.3\,{\rm TeV}$.

\begin{figure}[tbh]
\centering
\includegraphics[width=\linewidth]{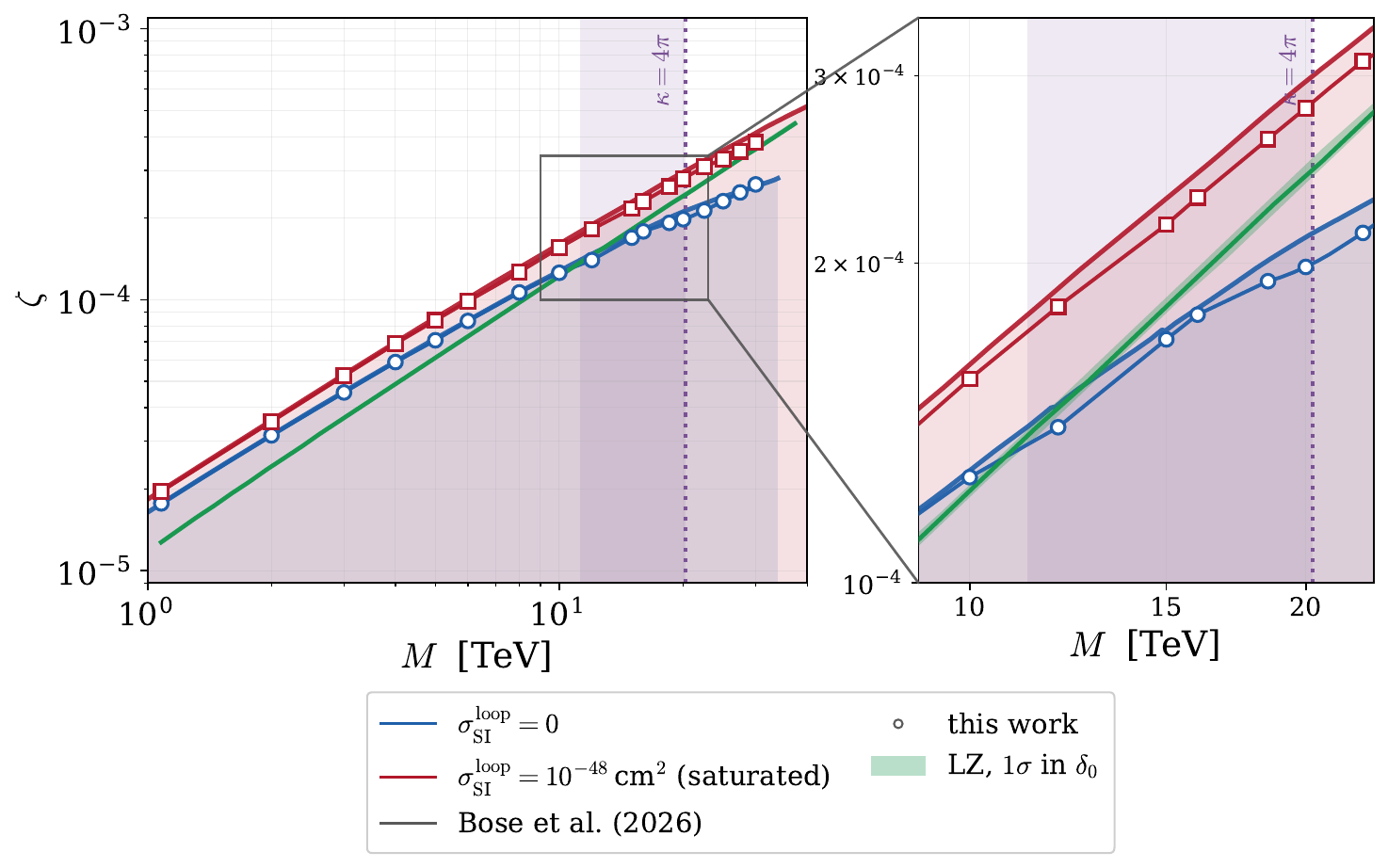}
\caption{The same constraints as Fig.~\ref{fig:Mdelta}, shown in the $(M,\zeta)$ plane,
where $\zeta$ is the $(H^\dagger\Phi)^2$ coupling that splits the neutral states,
$|\zeta|v^2 = 2M\delta_0 + \delta_0^2$. A fixed $\delta_0$ maps to a straight line here,
so the LZ-preferred splitting corresponds to $\zeta \sim 2\times10^{-4}$ at the candidate
masses: small, as required for $\zeta$ to be technically natural as the sole
$U(1)_\Phi$-breaking parameter. Curves, markers, shading and the $\kappa=4\pi$ line match
Fig.~\ref{fig:Mdelta}. The green band is the $1\sigma$ LZ-preferred range (best fit
$\delta_0\simeq368\,$keV, width $\sim15\,$keV). Right: magnification of the open window.
The LZ band crosses the $\sigma_{\rm SI}^{\rm loop}=0$ curve near $11\,$TeV and stays
below the saturated curve throughout the range shown.}
\label{fig:Mzeta}
\end{figure}

\section{Consistency of the Model}
\label{sec:consistency}

Having established a phenomenologically viable region of the parameter space, we examine the internal theoretical consistency of the model.

\subsection{Small $\sigma_{SI}$}

In the previous section, we derived the lower limit on dark matter mass $M>11.3$~TeV assuming $\sigma_{SI}=0$. According to \cite{Bose:2026ndd}, the lower limit on $\delta_0$ remains the same as long as $\sigma_{SI} < 10^{-51}$cm$^2$. Here we ask the question if it is reasonable to assume this small $\sigma_{SI}$.

Based on the estimate of the nucleon-Higgs Yukawa coupling \cite{Hoferichter:2017olk}, we find the tree-level contribution to be
\begin{align}
    \sigma_{SI}^{\rm tree} &= 6.29 \times 10^{-48} \eta^2 {\rm cm}^2 \left( \frac{11.3~{\rm TeV}}{M}\right)^{2}
    = 10^{-51}~{\rm cm}^2 \left( \frac{\eta}{0.0126} \right)^2 \left( \frac{11.3~{\rm TeV}}{M}\right)^{2}
    . \label{eq:sigmaSItree}
\end{align}
We assume $\eta$ at the relevant energy scale is small $|\eta| \lesssim 0.013$ so that the spin-independent elastic scattering is small enough for dark matter to stay in an extended orbit and the neutrino flux is suppressed \cite{Bose:2026ndd}. 

One may be concerned that the gauge interactions can induce loop-level $\sigma_{SI}$ independent of assumed tree-level couplings \cite{Klasen_2013}. The induced $\eta$ is easy to estimate using an effective potential, and we find $\Delta \eta = 0.0157$ using the renormalization scale $\mu=M$ to $m_W$. The non-Higgs diagram is known to be smaller \cite{Klasen_2013}. Therefore the assumed small $\eta$ is not upset by the one-loop gauge corrections. 

On the other hand, there is renormalization-group running of $\eta$ due to the large $\kappa$ coupling,
\begin{align}
    \mu \frac{d}{d\mu} \eta
    &= \frac{1}{8\pi^2} \kappa^2.
\end{align}
Given $\kappa \gtrsim 7$, $\Delta \eta \gtrsim 0.6 \ln \frac{\Lambda}{M}$, where $\Lambda$ is the UV cutoff, {\it e.g.}\/, Landau pole. The initial condition for $\eta$ needs to be tuned at a percent level to obtain the assumed small $|\eta| \lesssim 0.013$ at the dark matter scale.

\subsection{Perturbative unitarity}

\begin{figure}
    \centering
    \includegraphics[width=0.8\linewidth]{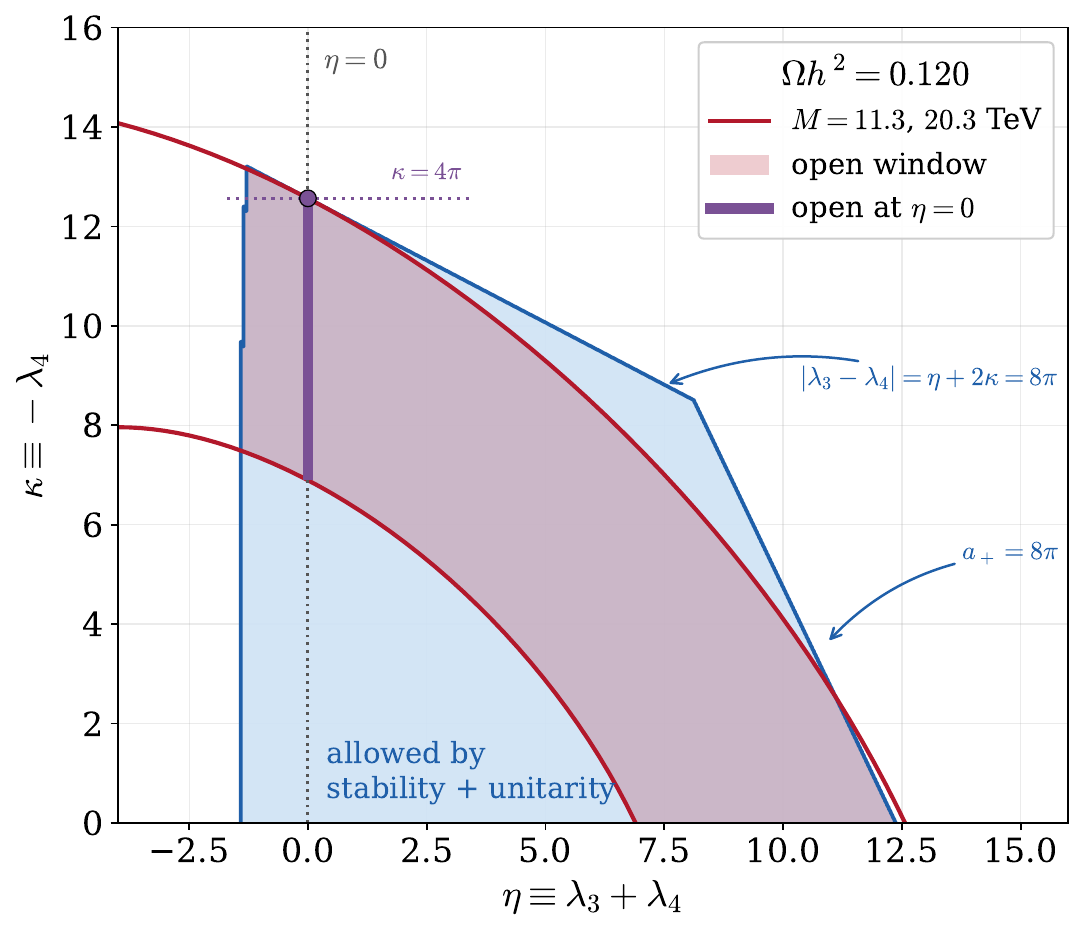}
    \caption{Parameter space consistent with perturbative unitarity and vacuum stability in
the $(\eta,\kappa)$ plane, with the phenomenologically viable region superimposed. The
pale blue region is allowed by boundedness from below together with
$|\Lambda|_{\rm max}<8\pi$, for some choice of $\lambda_\Phi$; its upper boundary is set
by $|\lambda_3-\lambda_4| = \eta+2\kappa = 8\pi$ at small $\eta$ and by the $a_+$
eigenvalue at large $\eta$. The two red curves are the thermal relic condition
$\Omega h^2 = 0.120$ at the edges of the open mass window, $M = 11.3$ and $20.3\,$TeV,
and the pink region between them is the part of the plane the model leaves open. The
dotted horizontal line is $\kappa = 4\pi$, which the relic condition reaches exactly at
the upper edge $M = 20.3\,$TeV. The heavy purple segment is the viable range at
$\eta = 0$, the limit in which the tree-level elastic cross section vanishes: there
$\kappa$ runs from 6.92 at $11.3\,$TeV to $4\pi$ at $20.3\,$TeV. Small $\eta$ is required
independently by the solar constraint.}
    \label{fig:eta_kappa_plane}
\end{figure}

Tree-level $2\to2$ scattering of the eight real scalars in $H$ and $\Phi$
block-diagonalises into channels of definite hypercharge and weak isospin, and
perturbative unitarity requires every eigenvalue $\Lambda_i$ of the resulting
$36$-dimensional matrix to satisfy $|\Lambda_i| < 8\pi$ --- equivalently
$|\Re e\, a_0| < \frac{1}{2}$ for the $s$-wave partial-wave amplitude
$a_0 = -\Lambda_i/16\pi$. The complete eigenvalue set for the general
two-Higgs-doublet potential, and a thorough discussion of the resulting constraints,
may be found in Ref.~\cite{Treesukrat:2024akz}. Only two of the eigenvalues can bind
when $\kappa$ is the largest coupling, and we quote those:
\begin{align}
    f &= \lambda_3-\lambda_4 = \eta+2\kappa,
    \label{eq:unitarity_f}
    \\
    a_\pm &= \frac{3}{2}\,(\lambda_H+\lambda_\Phi)
    \pm\sqrt{\frac{9}{4}\,(\lambda_H-\lambda_\Phi)^2+(2\eta+\kappa)^2}\ .
    \label{eq:unitarity_aplus}
\end{align}
Here $f$ is the amplitude in the $SU(2)$-singlet channel
$\frac{1}{\sqrt{2}}(\ket{H^+\Phi^0} - \ket{H^0\Phi^+})$, and it is $f$ that is
strongest in our case: with $\kappa$ the only large coupling $f\simeq2\kappa$, so
$|f|<8\pi$ gives
\begin{equation}
    |\kappa| < 4\pi
    \label{eq:kappa4pi}
\end{equation}
which we require in the remainder of the discussion. The $a_+$ eigenvalue grows with
$\lambda_\Phi$, and overtakes $f$ only once $\eta$ becomes comparable to $\kappa$ ---
far outside the region of interest, since the solar constraint already demands $\eta\lesssim10^{-2}$. We nonetheless draw
Fig.~\ref{fig:eta_kappa_plane} out to $\eta\sim\kappa$, so that both boundaries of the
allowed region are visible.

In addition to the unitarity limit, boundedness of the potential from
below requires
\begin{align}
    \lambda_H&>0,
    &
    \lambda_\Phi&>0,
    \\
    \eta+\kappa&>-\sqrt{\lambda_H\lambda_\Phi}\ ,
    &
    \eta-|\zeta|
    &>-\sqrt{\lambda_H\lambda_\Phi}\ .
\end{align}
The parameter space on the $(\eta,\kappa)$ plane consistent with the perturbative unitarity as well as the boundedness of the potential from below is shown in Fig.~\ref{fig:eta_kappa_plane}. It also superimposes the phenomenologically viable region of $M=11.3$--20.3~TeV (Fig.~\ref{fig:Mdelta}) and small $\eta$ as we will discuss later in the paper.

\subsection{Electroweak Constraint}

Given the large $\kappa$ that splits the mass of charged and neutral components, we need to see if it satisfies the electroweak constraints, in particular $T$ and $S$ parameters \cite{Peskin:1991sw}. We find that the constraints are easily satisfied.

The $T$-parameter from a split electroweak doublet scalar is \cite{Drees:1990dx}
\begin{align}
     \Delta T &= \frac{N_c}{16\pi m_W^2 \sin^2 \theta_W}
     \left[ m_1^2 + m_2^2 - \frac{2 m_1^2 m_2^2}{m_1^2-m_2^2} \ln \frac{m_1^2}{m_2^2} \right] .
\end{align}
For our case, $m_1^2 = m_{\phi^+}^2 = M^2 + \frac{1}{2} \kappa v^2$, $m_2^2 = m_{\phi^0}^2 = M^2$, and $N_c=1$. We find
\begin{align}
    \Delta T &= 0.0015 \left( \frac{\kappa}{6.92} \right)^2
    \left( \frac{\rm 11.3~TeV}{M} \right)^2.
\end{align}
On the other hand, 
\begin{align}
    \Delta S &= - \frac{N_c Y}{6\pi} \ln \frac{m_1^2}{m_2^2}
    = -0.000043\left( \frac{\kappa}{6.92} \right)
    \left( \frac{\rm 11.3~TeV}{M} \right)^2
\end{align}
They easily satisfy the experimental constraints $T=0.021 \pm 0.055$ and $S=0.008\pm 0.071$ \cite{ParticleDataGroup:2026mpi}. 

\subsection{Landau Pole}

\begin{figure}[t]
    \centering
    \includegraphics[width=0.8\linewidth]{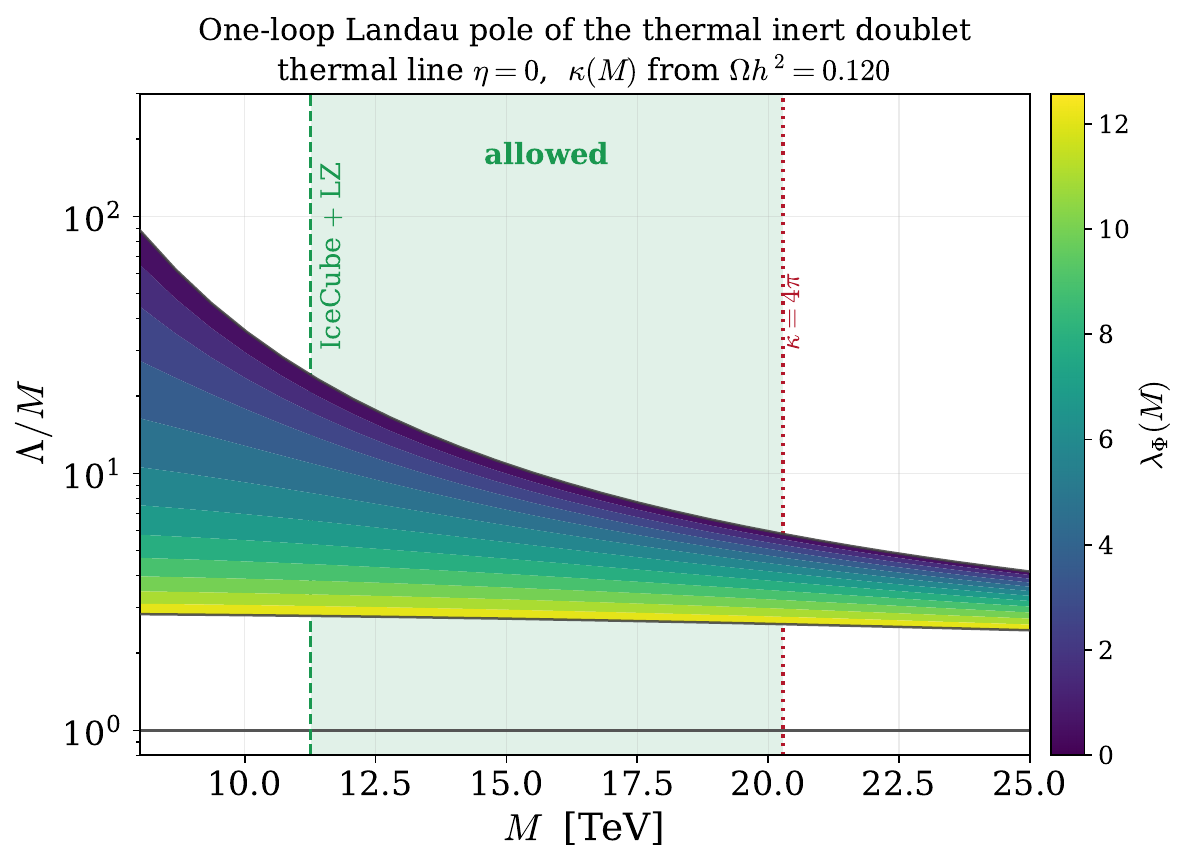}
    \caption{Validity range of the model: the ratio $\Lambda/M$ of the one-loop Landau pole to
the dark matter mass, along the thermal line $\eta=0$ with $\kappa(M)$ fixed by
$\Omega h^2 = 0.120$. $\Lambda$ is defined as the scale at which one of the quartic
couplings diverges, obtained by running the full one-loop system of
Ref.~\cite{Goudelis:2013uca}; the colour scale is the inert self-coupling
$\lambda_\Phi(M)$, the one quartic left free. The shaded region is the open mass window,
bounded below by the LZ and solar constraints at $11.3\,$TeV and above by
$\kappa<4\pi$ at $20.3\,$TeV. Across that window the pole sits between about $23\times M$
at the lower edge and $6\times M$ at the upper for $\lambda_\Phi\to0$, falling to
$\sim2.6\times M$ at $\lambda_\Phi = 4\pi$ --- roughly an order of magnitude of validity
range, narrowing as $\kappa$ grows with $M$. Note that the validity range is as large as that in the chiral Lagrangian $m_\pi/4\pi f_\pi\approx 0.12$. The model therefore needs a UV completion not
far above the dark matter mass, but remains a controlled effective theory throughout the
region of interest. }
    \label{fig:landau_pole}
\end{figure}

Given the large coupling $\kappa$, we need to know if there is a good validity range of the model before hitting the Landau pole. The one-loop renormalization group equations for the couplings in the potential were worked out in \cite{Goudelis:2013uca}. We computed the energy scale of the Landau pole defined by one of the couplings diverging as shown in Fig.~\ref{fig:landau_pole}. Yet we see that there is about an order of magnitude of the validity range for the phenomenologically interesting dark matter mass range. If there are additional Yukawa or gauge interactions, the window can be even bigger. Note that this is roughly the same as $m_\pi=140$~MeV compared to $4\pi f_\pi=1170$~MeV in chiral Lagrangian, and is much wider than $m_K=498$~MeV in the same context. It is clear that we need a UV completion of the model beyond the Landau pole. We do not see a problem with the model as an effective low-energy theory. 

The large coupling may be a consequence of compositeness of dark matter and Higgs, or other non-abelian gauge interactions that enhance the scalar coupling in the renormalization group evolution. We leave the discussion on an explicit UV completion to a future work.

\FloatBarrier
\section{Conclusion}
\label{sec:conclusion}

We have shown that the inert doublet dark matter can explain the purported direct detection event by the LZ experiment while evading the limit on high-energy solar neutrino flux by the IceCube experiment. The preferred dark matter mass is 11.3--20.3~TeV which is possible when a coupling $\kappa (H \Phi)(H \Phi)^*$ in the scalar potential is rather large but satisfies the perturbative unitarity $\kappa < 4\pi$. The small mass splitting $m_{\phi_3} - m_{\phi_0} \sim 370$~keV is technically natural in this model. The compatibility with the IceCube constraint requires the coupling $\eta (H^\dagger H)(\Phi^\dagger \Phi)$ to be small, $\eta \lesssim 0.02$. The model is internally consistent with constraints from perturbative unitarity, Landau pole, electroweak precision tests.

The most important near-future experimental effort is obviously more data from xenon-based direct detection experiments, LZ itself as well as XENONnT and PandaX-4T. The XLZD proposal \cite{XLZD:2024nsu} may have gained a stronger case. Proposed IceCube G2 would be uniquely suited for the indirect detection with a ten times larger volume. Cherenkov Telescope Array can look for gamma rays from galactic center and dwarf galaxies. Given the large $\kappa$, it should approximately correspond to the weak isospin of $O(10)$ and may be seen in the continuum \cite{Baumgart:2025dov}. The mass scale is unfortunately too high for 10~TeV muon collider or even FCC-hh given that $\Phi$ can be produced only by electroweak Drell--Yan processes. It calls for even more ambitious proposals. 

\acknowledgments
H.M. is supported by the NSF grant PHY-2515115, by the U.S. Department of Energy (DE-AC02-05CH11231), by the JSPS Grant-in-Aid for Scientific Research JP23K03382, MEXT Grant-in-Aid for Transformative Research Areas (A) 26H00401, 26A204, 26H00403, Hamamatsu Photonics, K.K., Tokyo Dome Corporation, and by the World Premier International Research Center Initiative (WPI) MEXT, Japan. B.N. is supported by the NSF grant PHY-2515115 and the UC Dissertation-Year Fellowship. The authors acknowledge the use of Claude Code (Anthropic) Sonnet 5 and Opus 5.5 for assistance in developing and debugging analysis scripts used in this work. The authors take full responsibility for the integrity and accuracy of the code and the scientific results presented herein.

\bibliographystyle{JHEP}
\bibliography{refs}

\end{document}